\documentclass[
 reprint,
superscriptaddress,
 amsmath,amssymb,
 aps, nofootinbib
]{revtex4-1}

\usepackage{graphicx}
\usepackage{dcolumn}
\usepackage{bm}
\usepackage{hyperref}
\usepackage[utf8]{inputenc}

\begin{document}

\preprint{preprint number}

\title{Scattering of massless scalar waves by magnetically charged black holes in Einstein-Yang-Mills-Higgs theory}

\author{Alexander Gu\ss mann}
 \email{alexander.gussmann@physik.uni-muenchen.de}
\affiliation{%
 Arnold Sommerfeld Center, Ludwig-Maximilians-Universit\"at, 80333 M\"unchen, Germany\\
}

\date{\today}

\begin{abstract}
The existence of the classical black hole solutions of the Einstein-Yang-Mills-Higgs equations with non-abelian Yang-Mills-Higgs hair, which we will also refer to as ``magnetic monopole black hole solutions", implies that not all classical stationary magnetically charged black holes can be uniquely described by their asymptotic characteristics. In fact, in a certain domain of parameters, there exist different spherically-symmetric, non-rotating and asymptotically-flat classical black hole solutions of the Einstein-Yang-Mills-Higgs equations which have the same ADM mass and the same magnetic charge but significantly different geometries in the near-horizon regions. (These are black hole solutions which are described by a Reissner-Nordstr\"om metric on the one hand and the hairy magnetic monopole black hole solutions which are described by a metric which is not of Reissner-Nordstr\"om form on the other hand.) One can experimentally distinguish such black holes with same asymptotic characteristics but different near-horizon geometries classically by probing the near-horizon regions of the black holes. We argue that one way to probe the near-horizon region of a black hole which allows to distinguish magnetically charged black holes with same asymptotic characteristics but different near-horizon geometries is by classical scattering of waves. Using the example of a minimally-coupled massless probe scalar field scattered by magnetically charged black holes which can be obtained as solutions of the Einstein-Yang-Mills-Higgs equations with a Higgs triplett and gauge group SU(2) in the limit of an infinite Higgs self-coupling constant we show how, in this case, the scattering cross sections differ for the magnetically charged black holes with different near-horizon geometries but same asymptotic characteristics. We find in particular that the characteristic glory peaks in the cross sections are located at different scattering angles.
\end{abstract}

\maketitle

\section{Introduction}

In the spirit of Wheeler and Ruffini \cite{Ruffini:1971bza} it has been conjectured a long time ago that classical stationary black holes can be completely characterized by parameters which are associated to a Gauss law. This ``no-hair" conjecture has been proven in the case of Einstein-Maxwell theory \cite{Israel:1967wq, *Robinson:1975bv, *Mazur:1982db, *heusler} and for several types of non-Maxwellian matter \cite{Bekenstein:1972ny, *Bekenstein:1971hc, *Bekenstein:1972ky, *Teitelboim:1972pk, *Teitelboim:1972ps, *Hartle:1971qq}. For a long time all known classical black hole solutions of the Einstein field equations did not contradict to this no hair conjecture. However, from the late 80s on, many classical black hole solutions of the Einstein field equations (with different matter sources) which were interpreted as black holes with classical primary hair (not associable to a Gauss law) have been discovered (see \cite{Volkov:1998cc, *Bizon:1994dh, *Bekenstein:1996pn, *Herdeiro:2015waa, *Volkov:2016ehx} for some reviews). Many of the asymptotically-flat and spherically-symmetric black hole solutions of this kind turned out to be dynamically unstable (in Lyapunov sense). To our knowledge, the only asymptotically-flat and spherically-symmetric black hole solutions of this kind which are known to be stable against spherically-symmetric linear perturbations are some of the black hole solutions where the matter source in the Einstein field equations is taken by an energy momentum tensor which corresponds to a Lagrangian which in flat spacetime allows for topological solitons as lowest energy configurations. These are first, some of the skyrmion black hole solutions of the Einstein-Skyrme equations \cite{Luckock:1986tr, *Droz:1991cx, *Bizon:1992gb, *Heusler:1992av, *Tamaki:2001wca, *Shiiki:2005pb, *Nielsen:2006gb, *Gudnason:2016kuu, *Adam:2016vzf} and second, some of the magnetic monopole black hole solutions of the Einstein-Yang-Mills-Higgs equations \cite{Ortiz:1991eu, Lee:1991vy, Breitenlohner:1991aa, Breitenlohner:1994di, Aichelburg:1992st}. Therefore, from the knowledge today, it seems that in this sense topology is a necessary condition for spherically-symmetric and asymptotically-flat classical hairy black holes to be dynamically stable on the linear level. All these asymptotically-flat and spherically-symmetric hairy classical black hole solutions which are known to be stable against spherically-symmetric linear perturbations have in common that they are known only in a domain of parameters in which the black hole event horizon is located inside of a characteristic length scale which is associated to the soliton.

In this work we focus on magnetic monopole black holes. These are classical stationary black hole solutions of the Einstein-Yang-Mills-Higgs equations in a certain domain of parameters which are described by metrics which are not of Reissner-Nordstr\"om form \cite{Lee:1991vy, Breitenlohner:1991aa, Breitenlohner:1994di, Aichelburg:1992st}. In the limit of an infinite Higgs self-coupling constant with all other parameters in the system kept fixed, it was proven analytically that such magnetic monopole black hole solutions exist globally \cite{Breitenlohner:1994di} and that there are black hole solutions of this kind which are stable against spherically-symmetric linear perturbations \cite{Aichelburg:1992st}. These are two reasons why it is particularly interesting to focus on this limit, in which the Higgs field becomes infinitely massive, although many numerical results are also known beyond that limit \cite{Lee:1991vy, Breitenlohner:1991aa, Breitenlohner:1997hm, *Gal'tsov:1997rz, *Breitenlohner:1997ud}. In this work we consider such magnetic monopole black holes which can be obtained as solutions of the Einstein-Yang-Mills-Higgs equations in the limit of an infinite Higgs self-coupling constant (keeping all the other parameters fixed). We will argue in the outlook that some results we obtain for this limiting case qualitatively apply also for magnetic monopole black holes obtained as solutions of the Einstein-Yang-Mills-Higgs equations with finite Higgs coupling constant.

On top of magnetic monopole black hole solutions there exist also Reissner-Nordstr\"om black hole solutions of the Einstein-Yang-Mills-Higgs equations. These Reissner-Nordstr\"om black holes exist as solutions of the Einstein-Yang-Mills-Higgs equations in a different parameter domain than the magnetic monopole black hole solutions. As shown in \cite{Lee:1991vy, Breitenlohner:1991aa, Breitenlohner:1994di, Aichelburg:1992st}, there is however an overlapp between the parameter domains of existence of magnetic monopole black hole and Reissner-Nordstr\"om black hole solutions of the Einstein-Yang-Mills-Higgs equations. In this overlapp, for a given magnetic charge and given ADM mass, there exist different black hole solutions of the Einstein-Yang-Mills-Higgs equations which have the same asymptotic characteristics but significantly different geometries in the near-horizon regions. That is the domain of parameters we are interested in this work.

The purpose of the present paper is to investigate how one can distinguish stationary classical magnetically charged black holes which have the same asymptotic characteristics but significantly different near-horizon geometries in scattering experiments of classical waves. For this purpose we study classical scattering cross sections of a massless minimally-coupled probe scalar field scattered by magnetically charged black holes. We focus on black holes which are obtained as solutions of the Einstein-Yang-Mills-Higgs equations with a Higgs tripplet and gauge group $SU(2)$ in the limit of an infinite Higgs self-coupling constant. To illustrate the most important points we perform a detailed numerical analysis for one ``working example" of such a magnetic monopole black hole. We then compare the obtained scattering cross sections with the analogous scattering cross sections of a magnetically charged Reissner-Nordstr\"om black hole which has the same asymptotic characteristics as the magnetic monopole black hole of our working example. Here we focus on the effect which results from the gravitational interactions of the probe scalar field with the black holes. (For this purpose we neglect possible non-gravitational interactions between the scalar field and the magnetic monopole.) We find that one can indeed distinguish these stationary black hole configurations which are described by the same set of asymptotic parameters but have different near-horizon geometries by such scattering experiments. The scattering cross sections of the probe scalar field are different for magnetically charged black holes with same asymptotic characteristics but different near-horizon geometries. In particular the characteristic glory peaks in the cross sections are located at different scattering angles. We will argue that this is not only the case in our working example which we present to illustrate these results numerically, but is a generic feature of all magnetic monopole black holes (obtained as solutions of the Einstein-Yang-Mills-Higgs equations in the limit of an infinite Higgs self-coupling constant) which are part of the parameter domain in which also Reissner-Nordstr\"om black holes with same asymptotic characteristics do exist. In the second part of \cite{Dvali:2016mur} analogous scattering cross sections were calculated for some skyrmion black holes.

Our results, that scattering cross sections of external waves scattered by black holes which have the same asymptotic characteristics but different near-horizon geometries are significantly different, can have important astrophysical implications. For example, in the case one has a black hole with given asymptotic characteristics in nature, it might be interesting to do experiments to study the geometry of the black hole in the regime close to its event horizon in order to find out if this black hole is on the one hand of Schwarzschild, Reissner-Nordstr\"om or Kerr form or if it on the other hand carries some hair. Our results imply that one way how this can in principle be done is by scattering of external waves. Proposals for different experiments with the same purpose often go under the name ``testing the no-hair hypothesis", see for example \cite{Johannsen:2016uoh, Cardoso:2016ryw} for recent reviews.

The rest of the paper is organized as follows: In section II we briefly review the aspects of magnetic monopole black holes in Einstein-Yang-Mills-Higgs theory which are most relevant for our analysis. We recapulate how magnetic monopole black holes differ from Reissner-Nordstr\"om black holes with same asymptotic characteristics in the part of the parameter domain where both magnetic monopole black holes and Reissner-Nordstr\"om black holes do exist as solutions of the Einstein-Yang-Mills-Higgs equations. In section III we calculate scattering cross-sections of a massless minimally-coupled probe scalar field scattered by a magnetic monopole black hole. We both use the glory approximation and a complete partial wave analysis to determine these cross sections for our working example. In section IV we compare these cross sections with the known cross sections of the same scalar field now scattered by a Reissner-Nordstr\"om black hole which has the same ADM mass and same charge as the magnetic monopole black hole considered in section III. In section V we conclude with a summary and an outlook.

For the Minkowski metric we use the signature $(+,-,-,-)$. We use units in which the speed of light is set to one but both the Newton constant $G_N$ and the Planck constant $\hbar$ are kept explicit.

\section{Magnetically charged black holes in Einstein-Yang-Mills-Higgs theory}

In this section we briefly review some aspects of the spherically-symmetric and asymptotically-flat classical magnetic monopole black hole solutions of the Einstein-Yang-Mills-Higgs equations described by a metric not of Reissner-Nordstr\"om form which can be interpreted as black holes inside 't Hooft-Polyakov magnetic monopoles. After some speculations about the possibility of such black hole solutions in \cite{Ortiz:1991eu}, these solutions have been discovered and studied for example in \cite{Lee:1991vy, Breitenlohner:1991aa, Breitenlohner:1994di, Aichelburg:1992st} and are often regarded as examples for black holes with classical primary ``non-abelian" hair \cite{Volkov:1998cc, *Bizon:1994dh, *Bekenstein:1996pn, *Herdeiro:2015waa, *Volkov:2016ehx}. We review the domain of parameters for which these black hole solutions have been found and compare this domain of parameters with the parameter domain in which Reissner-Nordstr\"om black holes exist as black hole solutions of the Einstein-Yang-Mills-Higgs equations. We work in Einstein-Yang-Mills-Higgs theory with a Higgs triplett and gauge group $SU(2)$.

The matter Lagrangian we consider is the Lagrangian of Einstein-Yang-Mills-Higgs theory which is given by
\begin{equation}
\mathcal{L} =-\frac{1}{4}F_{\mu \nu}^aF^{\mu \nu a} + \frac{1}{2}\mathcal{D}_\mu \phi^a \mathcal{D}^\mu \phi^a - \frac{\lambda}{2}\left(\phi^a \phi^a - v^2\right)^2\, .
\label{lagrangian}
\end{equation}
Here 
\begin{equation}
F_{\mu \nu}^a = \partial_\mu A_\nu^a-\partial_\nu A_\mu^a - e \epsilon_{abc}A_\mu^b A_\nu^c\, ,
\end{equation}
\begin{equation}
\mathcal{D}_\mu \phi^a = \partial_\mu \phi^a - e \epsilon_{abc}A_\mu^b \phi^c\, ,
\end{equation}
where Greek indices refer to spacetime indices and Latin indices refer to $SU(2)$ indices. $e$ is the Yang-Mills gauge coupling constant with dimensionality\footnote{As noted in the introduction, we work in units in which the speed of light is set to one but both the Newton constant $G_N$ and the Planck constant $\hbar$ are kept explicit.}
\begin{equation}
[e] = \frac{1}{\sqrt{\mathrm{[mass]}\mathrm{[length]}}}\, ,
\end{equation}
$\lambda$ is the Higgs coupling constant with dimensionality
\begin{equation}
[\lambda] = \frac{1}{\mathrm{[mass]}\mathrm{[length]}}
\end{equation}
and $v$ is the vacuum expectation value of the Higgs field $\phi^a$ with dimensionality
\begin{equation}
[v] = \sqrt{\frac{\mathrm{[mass]}}{\mathrm{[length]}}}\, .
\end{equation}

In flat spacetime the Lagrangian (\ref{lagrangian}) gives rise to spherically-symmetric solitonic solutions (``'t Hooft-Polyakov magnetic monopoles") \cite{'tHooft:1974qc, *Polyakov:1974ek} when using the ansatzes
\begin{equation}
\phi^a = v h(r) e_r^a, A_0 = 0, A_i^a = \epsilon_{iak}\left(\frac{1-u(r)}{er}\right)e_r^k
\label{ansatzgauge}
\end{equation}
and appropriate boundary conditions for the ansatz functions $h(r)$ and $u(r)$. (Here $e_r^k$ is the $k$-th component of the unit vector in radial direction.)

The characteristic mass of such a magnetic monopole is given by
\begin{equation}
M_m = \frac{v}{e}\, 
\end{equation}
and therefore the corresponding characteristic gravitational radius of a 't Hooft-Polyakov magnetic monopole is
\begin{equation}
L_g = 2 M_m G_N = 2 \frac{v}{e} G_N\, .
\label{lg}
\end{equation}
One characteristic length scale $L$ is given by the Compton wavelength of the gauge field
\begin{equation}
L = \frac{1}{ev}\, .
\label{l}
\end{equation}
The mass of the Higgs field is given by
\begin{equation}
M_H = \sqrt{\lambda} v \hbar
\label{mh}
\end{equation}
and the Compton wavelength of the Higgs field is therefore given by
\begin{equation}
L_C = \frac{1}{\sqrt{\lambda}v}\, .
\label{lc}
\end{equation}

For later convenience we introduce the dimensionless parameters $\alpha$ and $\beta$ defined as the ratios of the relevant length scales (\ref{lg}), (\ref{l}) and (\ref{lc}):
\begin{equation}
\alpha^2 \equiv 2 \pi \frac{L_g}{L} = 4 \pi v^2 G_N\, ,
\label{alpha}
\end{equation}
\begin{equation}
\beta \equiv \frac{L}{L_C} = \frac{\sqrt{\lambda}}{e}\, .
\label{beta}
\end{equation}

The Lagrangian (\ref{lagrangian}) can be coupled to gravity by using the corresponding energy momentum tensor, 
\begin{equation}
T_{\mu \nu} = \frac{2}{\sqrt{-g}}\frac{\delta \left(\sqrt{-g} \mathcal{L}\right)}{\delta g^{\mu \nu}} \,,
\label{energymomentum}
\end{equation}
as a source \cite{Bais:1975gu, *Cho:1975uz, *VanNieuwenhuizen:1975tc, Lee:1991vy, Breitenlohner:1991aa, Breitenlohner:1994di}. 
Using the ansatzes (\ref{ansatzgauge}) for the fields $A_\mu^a$ and $\phi^a$ and for the metric $g_{\mu \nu}$ the spherically-symmetric ansatz
\begin{equation}
ds^2 = N^2(r)t(r) dt^2 -t(r)^{-1}dr^2 - r^2d\Omega^2\, ,
\label{ansatzmetric}
\end{equation}
there are in total the four ansatz functions $u(r)$, $h(r)$, $N(r)$ and $t(r)$ for which the Einstein field equations\footnote{In the limit of an infinite Higgs coupling constant $\lambda$ (with all other paramters kept fixed) we provide the complete set of equations obtained by plugging the ansatzes (\ref{ansatzgauge}), (\ref{ansatzmetric}) and (\ref{ansatzmetric2}) into (\ref{einstein}) and the Euler-Lagrange equation for the matter field in the appendix. Beyond that limit one can find the equations for example in \cite{Breitenlohner:1991aa}.} 
\begin{equation}
G_{\mu \nu} = 8 \pi G_N T_{\mu \nu}
\label{einstein}
\end{equation}
and the field equations for the matter fields can be solved after choosing appropriate boundary conditions for $u(r)$, $h(r)$, $N(r)$ and $t(r)$. For convenience we define the ansatz function $M(r)$ as
\begin{equation}
t(r) = \left(1-\frac{2G_NM(r)}{r}\right)\, .
\label{ansatzmetric2}
\end{equation}


Using the boundary conditions
\begin{equation}
u(\infty) = 0, h(\infty) = 1, N(\infty) = 1\, ,
\end{equation}
non-trivial solutions of the field equations with event horizon (with a value $r_h \neq 0$ such that $2G_N M(r_h) = r_h$) for different choices of the other required boundary conditions (one boundary condition for $M(r)$, one more boundary condition for $h(r)$ and one more boundary condition for $u(r)$) have been found.\footnote{There are also solutions of the Einstein field equations (\ref{einstein}) without event horizon \cite{Bais:1975gu, *Cho:1975uz, *VanNieuwenhuizen:1975tc, Ortiz:1991eu, Lee:1991vy, Breitenlohner:1991aa} (gravitating 't Hooft-Polyakov magnetic monopoles). We however focus on the solutions with event horizon in this work.} These solutions are characterized by the boundary conditions of $M(r)$, $h(r)$ and $u(r)$ and by the parameters $\alpha$ and $\beta$ \cite{Breitenlohner:1991aa, Aichelburg:1992st}. The solutions fall in two clases: First, the following functions solve the field equations (\ref{einstein}):
\begin{equation}
u(r) = 0, h(r) = 1, M(r) = M - \frac{4 \pi}{2e^2 r}, N(r) = 1\, .
\end{equation}
Here $M$ is an arbitrary constant set by the boundary condition $M(\infty)$ which sets the ADM mass $M_{ADM}$. In this case, the metric is of Reissner-Nordstr\"om form. For
\begin{equation}
M^2 \geq \frac{2 \pi}{G_N e^2}
\label{massine}
\end{equation}
the Reissner-Nordstr\"om metric describes black holes (configurations with horizon(s) and without a naked singularity). Second, there exists a different type of black hole solutions of the field equations (``magnetic monopole black holes") described by a metric which is not of Reissner-Nordstr\"om form. Solutions of this type have only been found numerically in a certain domain of parameters. Indeed, for a given value of $\beta$ there exists a maximal value $\alpha_{max}(\beta)$ and this type of black hole solutions have only been found for
\begin{equation}
0 \leq \alpha \leq \alpha_{max}(\beta)\, .
\label{cond1}
\end{equation}
On top of that for given values of $\alpha$ and $\beta$, there exists a maximal possible black hole event horizon size $r_h^{max, \alpha, \beta}$ and magnetic monopole black holes have only been found as solutions of (\ref{einstein}) when the boundary conditions and  the parameters $\alpha$ and $\beta$ are chosen such that
\begin{equation}
0 \leq r_h \leq r_h^{max, \alpha, \beta}\, .
\label{cond2}
\end{equation}
From a physical point of view the knowledge of magnetic monopole black holes only within this domain of parameters implies on the one hand that the magnetic monopole of known magnetic monopole black hole solutions is such that it not itself becomes a black hole and on the other hand that the event horizon of known magnetic monopole black holes is always located inside of the Compton wavelength of the gauge field.

For a more detailed discussion of these black hole solutions as well as for details on the numerical methods used for obtaining them we refer to \cite{Lee:1991vy, Breitenlohner:1991aa, Breitenlohner:1994di, Aichelburg:1992st}. Here, we only want to emphasize that there is one particular limit in which different aspects of the magnetic monopole black hole solutions have been studied in great detail. This is the limit in which the mass of the Higgs field (\ref{mh}) is taken to infinity by taking the Higgs coupling constant $\lambda$ to infinity with all other parameters kept fixed. In this limit $\beta \rightarrow \infty$ and the Higgs field is frozen in its vacuum expectation value $h(r) = 1$. The Einstein field equations (\ref{einstein}) simplify in this limit allowing for several investigations which have not yet been worked out for magnetic monopole black holes with arbitrary values of $\beta$. In particular in \cite{Breitenlohner:1994di} there was suggested an analytical proof of global existence of magnetic monopole black hole solutions of (\ref{einstein}) in this limit and in \cite{Aichelburg:1992st} on the linearized level a whole stability analysis was performed for magnetic monopole black holes in this limit. Since magnetic monopole black hole solutions of (\ref{einstein}) are most studied and best understood in this limit, it is, although there are also some drawbacks of this limit \cite{Aichelburg:1992st}, in particular interesting to study magnetic monopole black holes obtained as solutions of (\ref{einstein}) in this limit. In the following we restrict to this limit.

\begin{figure}
\centering
\includegraphics[scale=0.8]{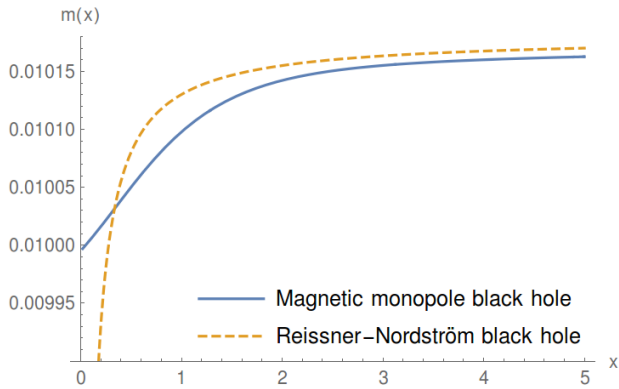}
\caption{$m(x)$ in the regime $x > x_h$ for the metric of a magnetic monopole black hole with $x_h = 0.02$, $m_{ADM} = 0.01018$, $\alpha = 0.01$ and for a Reissner-Nordstr\"om metric with $x_h = 0.01209$, $m_{ADM} = 0.01018$, $\alpha = 0.01$}
\label{fig:1}
\end{figure}

For the purpose of the present work it is important to mention that (in the above-mentioned limit) there is a part in the parameter space of black hole solutions of (\ref{einstein}) in which, for given asymptotic characteristics, there is a Reissner-Nordstr\"om black hole solution of (\ref{einstein}) described by a Reissner-Nordstr\"om metric which satisfies (\ref{massine}) and there is a magnetic monopole black hole solution of (\ref{einstein}) which satisfies (\ref{cond1}) and (\ref{cond2}) and which is described by a metric which is not of Reissner-Nordstr\"om form. A phase diagram of black hole solutions of (\ref{einstein}) can be found for example in \cite{Aichelburg:1992st}. This is the part of the parameter domain which we will consider. We choose one known \cite{Aichelburg:1992st} ``working example" in this part of the parameter domain: $\alpha = 0.01$, $\beta = \infty$, $x_h = 0.02$, $m_{ADM} = 0.01018$. In the next sections we illustrate our points by doing a numerical analysis for this working example. We will then argue that results we obtain for this example numerically, can qualitatively be applied for other magnetic monopole black holes (in the above-mentioned limit). For illustration, in Figure \ref{fig:1}, we plot the solution-function $m(x)$ for the magnetic monopole black hole solution with these parameters as well as the analogous function for the Reissner-Nordstr\"om black hole with same asymptotic characteristics as the magnetic monopole black hole of our working example\footnote{Here $x$ is a dimensionless variable defined as
\begin{equation}
x = e v r
\label{dimr}
\end{equation}
and $m(x)$ is a dimensionless function defined as
\begin{equation}
m(x)= e v G_N M(r)\, .
\label{dimmass}
\end{equation}
$x_h$ and $m_{ADM}$ is the corresponding dimensionless event horizon, dimensionless ADM mass respectively.},
\begin{equation}
m(x) = m_{ADM} - \frac{\alpha^2}{2x}\, .
\end{equation}
(We plot $m(x)$ only close to the event horizon in the regime $x \geq x_h$. Magnetic monopole black hole solutions in the regime $x < x_h$ are discussed for example in \cite{Breitenlohner:1997hm, *Gal'tsov:1997rz, *Breitenlohner:1997ud}.)

From the plot one can see that for our working example the event horizon size of the magnetic monopole black hole is bigger than the event horizon size of the Reissner-Nordstr\"om black hole with same asymptotic characteristics as the magnetic monopole black hole and that the shapes of the functions $m(x)$ significantly differ in the near-horizon regions. This is not only the case for our working example, but is a generic feature for all magnetic monopole black holes which are characterized by parameters out of the domain of parameters in which also Reissner-Nordstr\"om black hole solutions do exist as solutions of the Einstein field equations \cite{Aichelburg:1992st}. For $\alpha \rightarrow \alpha_{max}$ the metrics of magnetic monopole black holes approach the Reissner-Nordstr\"om metrics \cite{Aichelburg:1992st}.\footnote{In the case when the parameters are chosen such that only magnetic monopole black holes exist for such parameters, $r_h^{max, \alpha} \rightarrow 0$ for $\alpha \rightarrow \alpha_{max}$ \cite{Aichelburg:1992st}.}

\section{Classical scattering cross sections of a massless minimally-coupled probe scalar field scattered by a magnetic monopole black hole}

We take the magnetic monopole black hole described by the parameters of our working example and study classical scattering cross sections of a massless minimally-coupled probe scalar field $\Phi$ scattered by the magnetic monopole black hole. Both for simplicity and for later purposes which we will mention in section IV we neglect possible non-gravitational interactions between the magnetic monopole and $\Phi$. We argue later that the results we obtain in the case of our working example qualitatively apply also for all other magnetic monopole black holes (in the above-mentioned limit). Thus, the use of our working example is meant only to illustrate our results (which are qualitatively more generic) in one particular case.

The setup we are using can be generalized to waves of higher spin. In fact, instead of taking an external probe scalar field $\Phi$, one can perform the following analysis also for example with an external probe electromagnetic wave or an external probe gravitational wave scattered by the magnetic monopole black hole. In the present work we however restrict to the simplest case of an external probe scalar wave.

The motion of $\Phi$ in the background space-time of a magnetic monopole black hole can be described by the Klein-Gordon equation,
\begin{equation}
\Box_g \Phi = 0\, ,
\label{kleingordon}
\end{equation}
where $\Box_g$ is the d'Alambert operator in the space-time of a magnetic monopole black hole.

With the expansion
\begin{equation}
\Phi(t, r, \theta, \phi) = \sum_{lm} \frac{A_{Wl}(r)}{r} Y_{lm}(\theta, \phi)e^{-iWt}\, ,
\end{equation}
where $Y_{lm}(\theta, \phi)$ are the standard spherical harmonics, one can seperate (\ref{kleingordon}) into a radial part and an angular part. Using the metric (\ref{ansatzmetric}) and the dimensionless variable (\ref{dimr}), the radial part can be written as
\begin{equation}
\partial_{x^*}^2A_{wl}(x) + (w^2 - V_{eff}(x))A_{wl}(x) = 0\, ,
\label{radialpart}
\end{equation}
where $w$ is a dimensionless frequency defined as
\begin{equation}
w = W (e v)^{-1}\, .
\end{equation}
$x^*$ is defined as
\begin{equation}
\partial_{x^*} = N(x) t(x) \partial_x
\label{star}
\end{equation}
and the effective potential $V_{eff}(x)$ is given by
\begin{equation}
V_{eff}(x) = N^2(x) t(x) \frac{l(l+1)}{x^2} + \frac{N(x)}{x}t(x)\partial_x\left(N(x)t(x)\right)\, .
\label{effective}
\end{equation}
In this form the radial part of the Klein-Gordon equation (\ref{radialpart}) has the form of a Schr\"odinger equation. Cross sections can therefore be studied by using methods of one-dimensional quantum mechanical scattering theory.

Since we neglected non-gravitational interactions between $\Phi$ and the magnetic monopole, the matter fields $u(x)$ and $h(x)$ do not directly enter the equation (\ref{radialpart}) but only indirectly through their influence on the solution-functions $t(x)$ and $N(x)$ of (\ref{einstein}) which appear in (\ref{radialpart}) through (\ref{star}) and (\ref{effective}).

In the following subsections we study scattering cross sections for two monochromatic waves with the two frequencies $w = 100$ and $w = 300$ respectively. In subsection A, we use the glory approximation to determine the scattering cross sections of high frequency monochromatic waves for scattering angles $\theta \approx \pi$. In subsection B we use a partial wave analysis to obtain exact scattering cross sections for the two monochromatic waves with frequencies $w = 100$ and $w = 300$ respectively. We use the same methods which have been used frequently for example in studies of classical scattering cross sections in the case of waves scattered by Schwarzschild, Reissner-Nordstr\"om and Kerr black holes (see \cite{futterman} and references therein and \cite{Glampedakis:2001cx, *Dolan:2006vj, *Dolan:2008kf, *Crispino:2009xt, Crispino:2009ki, Dvali:2016mur} for some more recent studies).

\subsection{Geodesic motion and scattering cross sections using the glory approximation}

\begin{figure}
\centering
\includegraphics[scale=0.8]{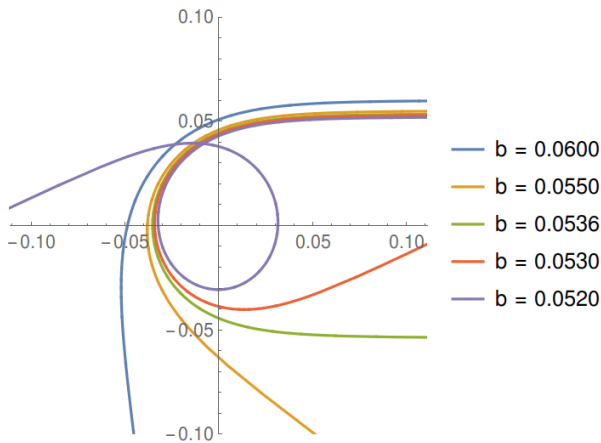}
\caption{Null geodesics with different impact parameters $b$ in the background of a magnetic monopole black hole with $x_h = 0.02$, $m_{ADM} = 0.01018$, $\alpha = 0.01$}
\label{fig:2}
\end{figure}

\begin{figure}
\centering
\includegraphics[scale=0.8]{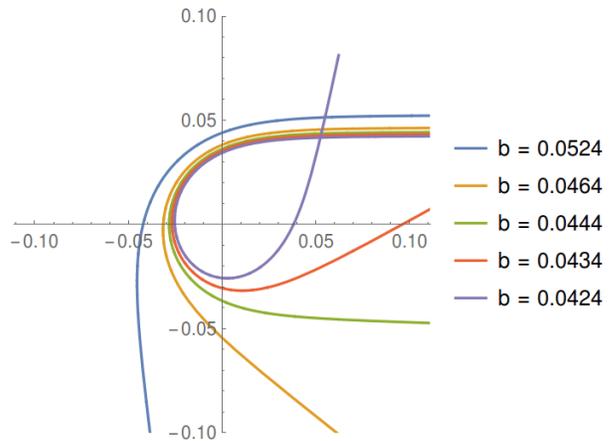}
\caption{Null geodesics with different impact parameters $b$ in the background of a Reissner-Nordstr\"om black hole with $x_h = 0.01209$, $m_{ADM} = 0.01018$, $\alpha = 0.01$}
\label{fig:3}
\end{figure}

Using a saddle point approximation it was shown for example in \cite{Handler:1980un, *matzner} that the differential scattering cross sections for high frequency scalar plane waves ($w \gg 1$) and for scattering angles $\theta \approx \pi$ can be obtained analytically as
\begin{equation}
\left(\frac{d\sigma}{d \Omega}\right)_{\theta \approx \pi} \approx 2 \pi w b_g^2 \left\lvert\frac{db}{d\theta}\right\rvert_{\theta = \pi} J_0^2\left(w b_g \mathrm{sin} \theta\right)\, .
\label{gloryapprox}
\end{equation}
Here $J_0$ is a Bessel function of first kind and $b_g$ is the impact parameter of the ray of a null geodesic in the background of a magnetic monopole black hole which is scattered in the same direction as it came from. We plot null geodesics for our working example of a magnetic monopole black hole for different impact parameters $b$ in Figure \ref{fig:2} and find $b_g = 0.0536$. We plot the corresponding differential scattering cross sections in Figure \ref{fig:4} and Figure \ref{fig:5} for the two cases $w=100$ and $w=300$.

For comparison, we plot null geodesics for the Reissner-Nordstr\"om black hole which has the same asymptotic characteristics as the magnetic monopole black hole of our working example in Figure \ref{fig:3}. For the null geodesics in this Reissner-Nordstr\"om black hole background we find $b_g^{(RN)} = 0.0444$. In particular,
\begin{equation}
b_g^{(RN)} < b_g
\label{impact}
\end{equation}
for our working example. By considering many examples we convinced ourselves that (\ref{impact}) is not only the case for our working example, but is a generic feature for all magnetic monopole black holes (obtained as solutions of the Einstein-Yang-Mills-Higgs equations in the limit of an infinite Higgs coupling constant) which are characterized by parameters within the domain of parameters in which also Reissner-Nordstr\"om black holes exist as solutions of the Einstein-Yang-Mills-Higgs equations.

\subsection{Scattering cross sections using a partial wave analysis}

We did a partial wave analysis to determine the scattering cross sections for our working example. In our working example, for all $l$, $V_{eff}(x^*) \rightarrow 0$ for $x^* \rightarrow \infty$. Therefore, one can write $A_{wl}(x^*)$ for $x^* \rightarrow \infty$ as
\begin{equation}
A_{wl}(x^*) = A^{(1)}_{wl} e^{-iwx^*} + A^{(2)}_{wl} e^{iwx^*}\, ,
\end{equation}
with the two complex coefficients $A^{(1)}_{wl}$ and $A^{(2)}_{wl}$. For $x^* \rightarrow -\infty$ we choose the boundary condition
\begin{equation}
A_{wl}(x^*) = A^{(3)}_{wl} e^{-iwx^*}\, ,
\end{equation}
where $A^{(3)}_{wl}$ is a complex coefficient which satisfies $|A^{(1)}_{wl}|^2+|A^{(2)}_{wl}|^2=|A^{(3)}_{wl}|^2$. With this boundary condition we describe a monochromatic wave (coming from infinity) which is purely ingoing (at the horizon).

The differential scattering cross section for a monochromatic scalar wave with frequency $w$ can be obtained as \cite{newton}
\begin{equation}
\frac{d\sigma}{d\Omega} = |h(\theta)|^2\, ,
\end{equation}
where
\begin{equation}
h(\theta) = \frac{1}{2iw}\sum_{l=0}^\infty (2l+1)\left(e^{2i\delta_l(w)}-1\right)P_l(\mathrm{cos} \theta)\, .
\label{sum}
\end{equation}
Here $P_l$ are the Legendre Polynomials and $\delta_l$ are ``phase shifts" which are defined as
\begin{equation}
e^{2i\delta_l(w)} = (-1)^{l+1}\frac{A^{(2)}_{wl}}{A^{(1)}_{wl}}\, .
\end{equation}

We determined the phase shifts $\delta_l$ from $l = 0$ for all $l$ up to a maximal value $l_{max}$. In order to calculate the sum (\ref{sum}) we used the ``method of reduced series" \cite{Yennie:1954zz} which was also used in \cite{Glampedakis:2001cx, *Dolan:2006vj, *Dolan:2008kf, *Crispino:2009xt, Crispino:2009ki, Dvali:2016mur}.

\begin{figure}
\centering
\includegraphics[scale=0.8]{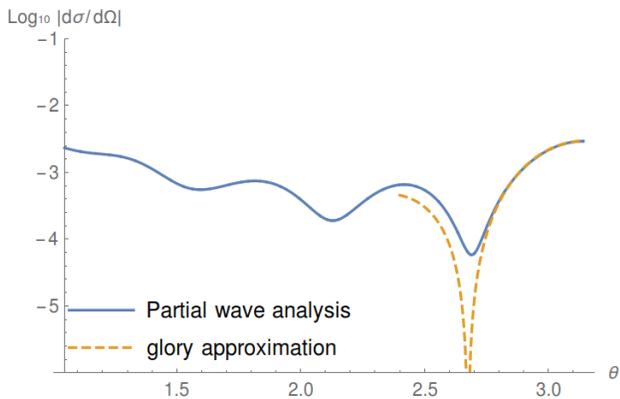}
\caption{Differential scattering cross section of a massless minimally-coupled probe scalar wave with frequency $w=100$ scattered by a magnetic monopole black hole with $x_h = 0.02$, $m_{ADM} = 0.01018$, $\alpha = 0.01$}
\label{fig:4}
\end{figure}

\begin{figure}
\centering
\includegraphics[scale=0.8]{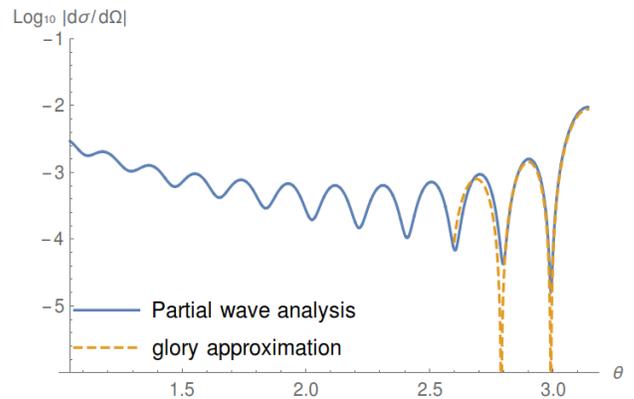}
\caption{Differential scattering cross section of a massless minimally-coupled probe scalar wave with frequency $w=300$ scattered by a magnetic monopole black hole with $x_h = 0.02$, $m_{ADM} = 0.01018$, $\alpha = 0.01$}
\label{fig:5}
\end{figure}

We plot the obtained differential scattering cross sections in Figure \ref{fig:4} and Figure \ref{fig:5}.

\section{Comparison of the scattering cross sections of magnetic monopole black holes and Reissner-Nordstr\"om black holes with same asymptotic characteristics}

We compare the differential scattering cross sections of the massless minimally-coupled probe scalar field scattered by magnetic monopole black holes with the known \cite{Crispino:2009ki} differential scattering cross sections of the same scalar field now scattered by Reissner-Nordstr\"om black holes which have the same asymptotic characteristics as the magnetic monopole black holes. Here we want to focus on the effects caused by the different metrics of these two classes of black hole solutions of (\ref{einstein}) with same asymptotic characteristics. That is one reason why in our analysis in section III we neglected possible non-gravitational interactions between the probe scalar field and the magnetic monopole which in principle can also lead to differences in the cross sections which we do not capture in our analysis.

\begin{figure}
\centering
\includegraphics[scale=0.8]{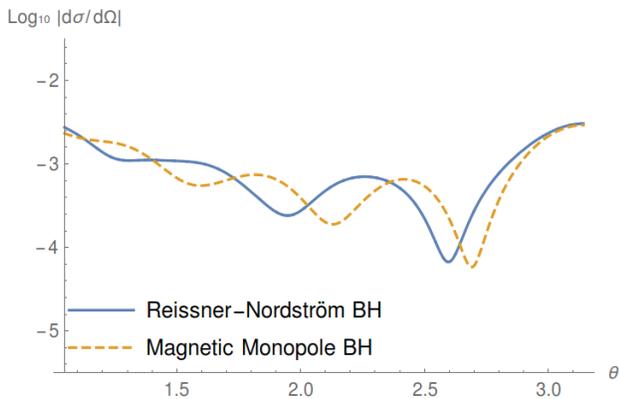}
\caption{Differential scattering cross section of a massless minimally-coupled probe scalar wave with frequency $w=100$ scattered by a magnetic monopole black hole with $x_h = 0.02$, $m_{ADM} = 0.01018$, $\alpha = 0.01$ and differential scattering cross section of the same scalar wave scattered by a Reissner-Nordstr\"om black hole with same asymptotic characteristics as the magnetic monopole black hole}
\label{fig:6}
\end{figure}

\begin{figure}
\centering
\includegraphics[scale=0.8]{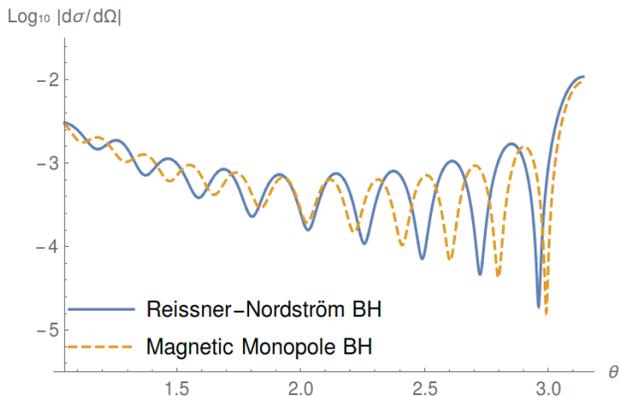}
\caption{Differential scattering cross section of a massless minimally-coupled probe scalar wave with frequency $w=300$ scattered by a magnetic monopole black hole with $x_h = 0.02$, $m_{ADM} = 0.01018$, $\alpha = 0.01$ and differential scattering cross section of the same scalar wave scattered by a Reissner-Nordstr\"om black hole with same asymptotic characteristics as the magnetic monopole black hole}
\label{fig:7}
\end{figure}

We plot the results for our working example in Figure \ref{fig:6} and Figure \ref{fig:7} for the two frequencies $w = 100$, $w = 300$ respectively. One can see from these figures that in the case of our working example the differential scattering cross sections are different for the black holes with same asymptotic characteristics but different near-horizon geometries. In particular, the characteristic ``glory" peaks in the cross sections of the scalar field scattered by magnetic monopole black holes are located at bigger scattering angles than the analogous peaks in the differential scattering cross sections of the same scalar field scattered by a Reissner-Nordstr\"om black hole with same asymptotic characteristics as the magnetic monopole black hole.

The result that the glory peaks are located at bigger scattering angles is not only true for our working example, but is a generic feature of all differential scattering cross sections caused by a minimally coupled massless probe scalar field scattered by any magnetic monopole black hole (obtained as solution of the Einstein-Yang-Mills-Higgs equations in the limit of an infinite Higgs coupling constant) which is characterized by parameters within the domain of parameters in which also Reissner-Nordstr\"om black holes do exist as solutions of the Einstein-Yang-Mills-Higgs equations. For the glory peaks located at scattering angles $\theta \approx \pi$ this is a direct consequence of (\ref{gloryapprox}) and the fact that (\ref{impact}) is generic for all such magnetic monopole black holes.

Since for $\alpha \rightarrow \alpha_{max}$ the metric of a magnetic monopole black hole approaches the Reissner-Nordstr\"om metric \cite{Aichelburg:1992st}, the differential scattering cross sections of a wave scattered by a magnetic monopole black hole becomes indistinguishable from the scattering cross section of the same wave scattered by a Reissner-Nordstr\"om black hole with the same asymptotic characteristics as the magnetic monopole black hole when $\alpha \rightarrow \alpha_{max}$.

\section{Outlook}

In this work we studied classical differential scattering cross sections of a massless minimally coupled probe scalar field scattered by magnetic monopole black holes which can be obtained as solutions of the Einstein-Yang-MIlls-Higgs equations with a Higgs triplett and gauge group $SU(2)$ in the limit of an infinite Higgs self-coupling constant. We compared these differential scattering cross sections with the analogous scattering cross sections of the same scalar field scattered by Reissner-Nordstr\"om black holes which have the same asymptotic characteristics as the magnetic monopole black holes but different near-horizon geometries. We found that the cross sections are different, in particular the characteristic glory peaks are located at different scattering angles. This implies that one can use scalar waves to probe near-horizon regions of black holes in order to experimentally distinguish magnetically charged black holes with same asymptotic characteristics but different near-horizon geometries. Similar results were obtained in the second part of \cite{Dvali:2016mur} for some skyrmion black holes. To illustrate our points numerically, we made use of one particular working example and we didn't take into account possible non-gravitational interactions between the probe scalar field and the 't Hooft-Polyakov magnetic monopole. (Taking such possible non-gravitational interactions into account requires further study.) We argued qualitatively how the results obtained for this example can be applied for arbitrary magnetic monopole black holes with infinite Higgs coupling constant.

For a more general (numerical) study which goes beyond the limit of an infinite Higgs coupling constant several changes would be necessary: Using a magnetic monopole black hole which is characterized by a finite Higgs mass requires, as in the limiting case considered in this work, a magnetic monopole black hole solution of the field equations (\ref{einstein}). In contrast to the case considered in this work where the Higgs field is frozen in its vacuum expectation value $h(r) = 1$, in this general case there is a set of four coupled differential equations which have to be solved with appropriate boundary values for the ansatz functions (instead of the three equations (\ref{app1}), (\ref{app2}) and (\ref{app3}) in the case we considered in this work). These equations and solutions for the general case are well-known and can be found for example in \cite{Lee:1991vy, Breitenlohner:1991aa, Breitenlohner:1994di}. In contrast to the limiting case used in this work there exist in this general case different branches of magnetic monopole black hole solutions and at most magnetic monopole black holes of one of these branches can be stable against perturbations \cite{Aichelburg:1992st}. It is often expected that in this general case the ``lower branch" of magnetic monpole black hole solutions is stable against perturbations whereas the ``upper branch" is not \cite{Aichelburg:1992st}. Therefore, in a quantitative analysis, it would be most interesting to investigate magnetic monopole black hole solutions from the lower branch.

An interesting question is if one can make qualitative statements about such cases with finite Higgs coupling constant without performing a complete quantitative numerical analysis. And indeed, in view of the analysis done in this work, we expect that such qualitative statements can be made. Since the main qualitative result of this work - the result that the characteristic glory peaks in the differential scattering cross sections of probe scalar fields scattered by magnetic monopole black holes (obtained in the limit of an infinite Higgs coupling constant) are located at bigger scattering angles than the peaks in the analogous scattering cross sections of a Reissner-Nordstr\"om black hole with same asymptotic characteristics - is a direct consequence of (\ref{impact}), the same qualitative result would apply for magnetic monopole black holes with finite Higgs coupling constant if and only if (\ref{impact}) holds also for those black holes. Since (\ref{impact}) is a consequence of the difference of the metrics of a Reissner-Nordstr\"om black hole and a magnetic monopole black hole in the near-horizon region, in particular of the difference of the solution mass functions $m(x)$ in the near-horizon region, we expect that in the case when the difference of the solution mass functions of magnetic monopole black holes and Reissner-Nordstr\"om black holes with same asymptotic characteristics is in the case of a finite Higgs coupling constant qualitatively the same as in the case of an infinite Higgs coupling constant, also the results in the differential scattering cross sections are qualitatively the same. Since to our knowledge this is indeed the case, at least for a large class of magnetic monopole black holes with finite Higgs coupling constant \cite{Breitenlohner:1991aa}, we expect that also for such magnetic monopole black holes the characteristic glory peaks in the scattering cross sections of a probe scalar field are located at bigger scattering angles when compared to the peaks in the analogous scattering cross sections caused by a Reissner-Nordstr\"om black hole with same asymptotic characteristics.

This work, as well as the similar analysis in the second part of \cite{Dvali:2016mur} for skyrmion black holes, opens up many other new possible directions for further studies. First, it might be interesting (instead of considering a massless minimally-coupled scalar field) to consider different kinds of waves, for example waves with higher spins, and to do a similar analysis for such different kind of waves. Second, one can study absorption cross sections of waves scattered by magnetic monopole black holes or skyrmion black holes. (Such an analysis would be similar to the analysis done in \cite{Macedo:2015ikq} where absorption cross sections of dirty black holes - also parametrized by a metric of the form (\ref{ansatzmetric}) - were studied.) Third, it might be very interesting to compare scattering cross sections caused by magnetically charged black hole solutions in Einstein-Yang-Mills-Higgs theory with scattering cross sections caused by Bardeen black holes which have been interpreted as magnetically charged black holes of some ``non-linear electrodynamics" coupled to gravity \cite{AyonBeato:2000zs}. For the case of Bardeen black holes a scattering analysis analogous to the one performed in this work for magnetic monopole black holes in Einstein-Yang-Mills-Higgs theory was done recently \cite{Macedo:2015qma, *Macedo:2016yyo} and it might be very interesting to compare the analysis done there with the anlaysis of the present work. Fourth, one might not only consider magnetic monopole black holes or skyrmion black holes but also black holes with different kind of classical hair such as the skyrmion magnetic monopole black holes \cite{Moss:2000hf} or the recently discovered Kerr black holes with scalar hair \cite{Herdeiro:2014goa, *Herdeiro:2015gia}. Such an analysis might have important astrophysical consequences for testing the no-hair conjecture experimentally \cite{Johannsen:2016uoh, Cardoso:2016ryw}. Fifth, one might study the scattering cross sections (which we studied in this work only classically) also quantum mechanically. Indeed, quantum effects might be important in particular since for realistic parameters both the magnetic monopole black holes and the skyrmion black holes are known as solutions of the Einstein field equations only in a domain of parameters in which these hairy black holes are microscopically small.

\begin{acknowledgments}
I want to thank Gia Dvali for many discussions about classical and quantum black hole hair. I want to thank C. F. B. Macedo, L. C. B. Crispino and E. S. de Oliveira for pointing me out their resent works on scattering cross sections of scalar waves in the background of a Bardeen black hole. This work was supported by the DFG cluster of excellence EXC 153 ``Origin and Structure of the Universe" and in part by the Humboldt Foundation.
\end{acknowledgments}

\appendix*

\section{Field equations in Einstein-Yang-Mills-Higgs theory}

In this appendix we provide the field equations for the gravitational field obtained by plugging the ansatzes (\ref{ansatzgauge}), (\ref{ansatzmetric}), (\ref{ansatzmetric2}) in the Einstein field equations (\ref{einstein}) and the Euler Lagrange equations for the matter fields in the limit of an infinitelly heavy Higgs field which is frozen in its vacuum expectation value $h(r) = 1$. We used these equations to calculate the scattering cross sections. The equations can also be found for example in \cite{Aichelburg:1992st} with slightly different conventions.

We use the dimensionless quantities (\ref{alpha}), (\ref{dimr}) and (\ref{dimmass}).
Plugging in the ansatzes (\ref{ansatzgauge}), (\ref{ansatzmetric}), (\ref{ansatzmetric2}) in the Einstein field equations (\ref{einstein}), it turns out that all the independent non-vanishing components of (\ref{einstein}) can be written in terms of the following equations:
\begin{equation}
\partial_x N(x) = 2 \alpha^2 \frac{N(x)}{x} (\partial_x u(x))^2
\label{app1}
\end{equation}
\begin{equation}
\partial_x m(x) = \alpha^2\left((\partial_x u(x))^2t(x)+\frac{(1-u(x)^2)^2}{2x^2}+u(x)^2\right)
\label{app2}
\end{equation}
\begin{equation}
\partial_x\left(\partial_x u(x)N(x)t(x)\right) = \frac{N(x)u(x)}{x^2}\left(x^2 - (1-u(x)^2)\right)
\label{app3}
\end{equation}
where the last equation can also be obtained from the matter field equation.

\end{document}